\documentclass[aip,pop,preprint]{revtex4-1}
\usepackage[utf8]{inputenc}
\usepackage{amsmath,amssymb,siunitx}
\usepackage{graphicx,float,dcolumn,bm}
\usepackage{hyperref,comment}
\usepackage{natbib}
\usepackage{silence}
\begin{document}

\title[Feasibility and Performance of the SZP: 1-D FLASH and MACH2]{Feasibility and Performance of the Staged Z-Pinch: A One-dimensional Study with FLASH and MACH2}

\author{E.~C.~Hansen}
\email{ehansen@pas.rochester.edu}
\affiliation{University of Rochester, Rochester, New York 14627, USA}
\author{F.~Garcia-Rubio}
\affiliation{University of Rochester, Rochester, New York 14627, USA}
\author{M.~B.~P.~Adams}
\affiliation{University of Rochester, Rochester, New York 14627, USA}
\author{M.~Fatenejad}
\affiliation{University of Rochester, Rochester, New York 14627, USA}
\author{K.~Moczulski}
\affiliation{University of Rochester, Rochester, New York 14627, USA}
\author{P.~Ney}
\affiliation{Magneto-Inertial Fusion Technology Inc., Tustin, California 92780, USA}
\author{H.~U.~Rahman}
\affiliation{Magneto-Inertial Fusion Technology Inc., Tustin, California 92780, USA}
\author{A.~C.~Reyes}
\affiliation{University of Rochester, Rochester, New York 14627, USA}
\author{E.~Ruskov}
\affiliation{Magneto-Inertial Fusion Technology Inc., Tustin, California 92780, USA}
\author{V.~Tranchant}
\affiliation{University of Rochester, Rochester, New York 14627, USA}
\author{P.~Tzeferacos}
\affiliation{University of Rochester, Rochester, New York 14627, USA}

\date{\today}

\begin{abstract}
Z-pinch platforms constitute a promising pathway to fusion energy research.
Here, we present a one-dimensional numerical study of the staged Z-pinch (SZP) concept using the \emph{FLASH} and \emph{MACH2} codes.
We discuss the verification of the codes using two analytical benchmarks that include Z-pinch-relevant physics, building confidence on the codes' ability to model such experiments.
Then, \emph{FLASH} is used to simulate two different SZP configurations: a xenon gas-puff liner (SZP1*) and a silver solid liner (SZP2).
The SZP2 results are compared against previously published \emph{MACH2} results, and a new code-to-code comparison on SZP1* is presented.
Using an ideal equation of state and analytical transport coefficients, \emph{FLASH} yields a fuel convergence ratio (CR) of approximately $39$ and a mass-averaged fuel ion temperature slightly below 1 keV for the SZP2 scheme, significantly lower than the full-physics \emph{MACH2} prediction.
For the new SZP1* configuration, full-physics \emph{FLASH} simulations furnish large and inherently unstable CRs ($>$ 300), but achieve fuel ion temperatures of many keV.
While \emph{MACH2} also predicts high temperatures, the fuel stagnates at a smaller CR.
The integrated code-to-code comparison reveals how magnetic insulation, heat conduction, and radiation transport affect platform performance and the feasibility of the SZP concept.
\end{abstract}

\maketitle

\section{\label{sec:intro}Introduction}
The Z-pinch concept is fundamentally a cylindrical plasma implosion onto the symmetry axis by a $\mathbf{J} \times \mathbf{B}$ force provided by a current pulse.
There are many variations on the target plasma, such as foils, wire arrays, jets, gas-puffs, pre-filled cylinders, or combinations thereof. \cite{Haines2000}
Furthermore, additional materials can be used as liners to assist the implosion, with yet more variations on how the liner is created and which material is used.
When such a system is driven by modern pulsed-power drivers, the current pinching the target can reach many MA, leading to plasmas that achieve keV temperatures at near-solid densities.
These plasmas are of interest to the fusion community and are useful scientific platforms for atomic physics, radiation transport, and laboratory astrophysics studies. \cite{Matzen1997,Haines2000}

The Z Machine at Sandia National Laboratories in Albuquerque (SNL) is the most powerful pulsed-power device in the world, providing up to 30 MA of peak current to a Z-pinch target. \cite{Sinars2020}
In recent years, the Magnetized Liner Inertial Fusion (MagLIF) concept has been a focus of research and development at SNL.
MagLIF is a specific type of Z-pinch that utilizes an externally applied axial magnetic field to reduce thermal conduction losses and an on-axis laser to preheat the fuel (typically close to 100 eV), which reduces the implosion velocity required to reach ignition temperatures. \cite{Slutz2010,Slutz2012}
The axial magnetic field is initially 10-20 T but is compressed to much larger values, which my help confine alpha particles when deuterium-tritium fuel is used.
MagLIF typically uses an aluminum or beryllium liner to compress a deuterium target and require sufficient liner thickness to avoid significant degradation due to the Rayleigh-Taylor instability.

The staged Z-pinch (SZP) is an alternative fusion concept in which energy is transferred to the target plasma in stages.
The SZP name was first used for a configuration with an on-axis cryogenic deuterium fiber (i.e., target) compressed by an argon or krypton liner. \cite{Rahman1995}
A current pre-pulse through the fiber would create the target plasma and pre-magnetize the liner, and a subsequent main Z-pinch current pulse would implode the liner.  
A theorized benefit of the SZP is the control and mitigation of the magneto-Rayleigh Taylor (MRT) instability at the fuel/liner interface \cite{Rahman2004}, however this point is beyond the scope of this work and will be addressed in future publications.
Current SZP configurations typically employ a gas fill for the target load and high atomic number liners (gas-puff liners \cite{Ney2001,Rahman2009Plasma} or solid liners \cite{Wessel2015,Wessel2016}).
The working hypothesis is that a high atomic number liner will radiate more efficiently, and the resulting colder liner will allow more magnetic diffusion towards the fuel/liner interface.
This would in turn result in a stronger magnetic pressure on the target plasma and potentially reduce thermal conduction losses.

It is well known that fusion output is severely inhibited when high atomic number impurities are mixed into the fuel plasma, because this increases radiative losses (i.e., reduces fuel temperatures).
Therefore, the high atomic number liners used by the SZP concept will only perform well if the fuel/liner interface remains relatively stable during the implosion.
The magnetic, thermal, and radiation transport properties of the system become crucially important as they can all affect the time scales of the implosion and fuel heating and stability of the fuel/liner interface.
We focus on the transport properties and their effects in the simulations presented in this work, but, as previously mentioned, we do not include a stability analysis as these simulations are one-dimensional.

\begin{figure}[H]
    \centering
    \includegraphics[width=0.8\linewidth]{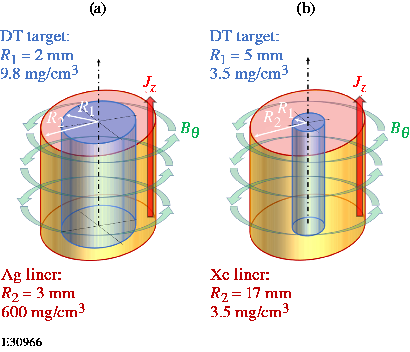}
    \caption{Schematics showing the SZP2 (a) and SZP1* (b) configurations. The SZP2 configuration uses a DT fuel and a silver solid liner, whereas SZP1* uses DT fuel and a xenon gass-puff liner.}
    \label{fig:schematics}
\end{figure}

In this paper, we have modeled two different SZP configurations: a new xenon gas-puff liner (SZP1*) with different initial conditions as compared to the original xenon gas-puff liner (SZP1 \cite{Rahman2009Plasma}), and the original silver solid liner setup (SZP2 \cite{Wessel2015}).
Fig.~\ref{fig:schematics} shows schematics of the SZP2 and SZP1* configurations with approximate dimensions.

The SZP concept has been subject to strong criticism \cite{Lindemuth2018,Lindemuth2021}, some of which has been addressed previously. \cite{Ruskov2020,Ruskov2021a,Ruskov2021b}
Most of the criticism has been aimed at the interpretation of key shock physics and the calculations behind the fusion energy output.
The \emph{FLASH} code can now contribute to this debate courtesy of our on-going collaboration between the Flash Center for Computational Science at the University of Rochester and MIFTI, made possible by funding from the ARPA-E BETHE program.
For the present work, we focus on specific physics and code-to-code comparisons, and we exclude calculations and discussion of fusion yield and energy production, in part, because \emph{FLASH} does not have this capability.

This paper is written with several goals in mind:
(1) to introduce \emph{FLASH}'s new capability of modeling Z-pinches,
(2) to further verify both \emph{FLASH} and \emph{MACH2} against analytical test problems and with direct code-to-code comparisons of SZP simulations,
(3) to provide a new SZP configuration (SZP1*) for additional verification and future experimental validation, and
(4) to shed more light on some of the previously published work by presenting SZP2 results from \emph{FLASH}. 

The structure of the paper is as follows:
in Section~\ref{sec:methods} we describe the two codes used in this work, \emph{FLASH} and \emph{MACH2}.
Then in Section~\ref{sec:tests}, we present results from two analytical test problems with SZP-relevant physics: a radiative shock problem and the Noh cylindrical implosion problem.
We show \emph{FLASH} results from an ideal equation of state (EOS) silver liner (SZP2) model in Section~\ref{sec:SZP2}, and we briefly discuss how it compares to the originally published \emph{MACH2} SZP2 results.
In Section~\ref{sec:SZP1}, we include more realistic EOS tables and physics to present both \emph{FLASH} and \emph{MACH2} simulation results of a xenon gas-puff liner configuration (SZP1*).
Lastly, we conclude our findings in Section~\ref{sec:conc}.

\section{\label{sec:methods}Numerical Methods}
\emph{FLASH} \cite{Fryxell2000} is a publicly-available, parallel, multi-physics, adaptive mesh refinement (AMR), finite-volume Eulerian hydrodynamics and MHD code, developed at the University of Rochester by the Flash Center for Computational Science (for more information on the \emph{FLASH} code, visit: \protect\url{https://flash.rochester.edu}).
\emph{FLASH} scales well to over a 100,000 processors, and uses a variety of parallelization techniques like domain decomposition, mesh replication, and threading, to optimally utilize hardware resources.
The \emph{FLASH} code has a world-wide user base of more than 4,350 scientists, and more than 1,300 papers have been published using the code to model problems in a wide range of disciplines, including plasma astrophysics, combustion, fluid dynamics, high energy density physics (HEDP), and fusion energy.

Over the past decade and under the auspices of the U.S. DOE NNSA, the Flash Center has added in \emph{FLASH} extensive HEDP and extended-MHD capabilities \cite{Tzeferacos2015} that make it an ideal tool for the multi-physics modeling of the SZP platform.
These include multiple state-of-the art hydrodynamic and MHD shock-capturing solvers,~\cite{Lee2013}, three-temperature extensions~\cite{Tzeferacos2015} with anisotropic thermal conduction that utilizes high-fidelity magnetized heat transport coefficients,~\cite{JiHeld2013} heat exchange, multi-group radiation diffusion, tabulated multi-material EOS and opacities, laser energy deposition, circuit models,~\cite{McBride2010} and numerous synthetic diagnostics \cite{Tzeferacos2017}. 
\emph{FLASH}'s newest algorithmic developments include a complete generalized Ohm's law that incorporates all extended MHD terms of the Braginskii formulation \cite{Braginskii1965}.
The new extended MHD capabilities are integrated with state-of-the-art transport coefficients,~\cite{Davies2021}  developed with support from the BETHE program.

The \emph{FLASH} code and its capabilities have been validated through benchmarks and code-to-code comparisons,~\cite{Fatenejad2013, Orban2013, Orban2022} as well as through direct application to numerous plasma physics experiments,~\cite{Meinecke2014, Meinecke2015, Li2016, Tzeferacos2018, Chen2020, Bott2021, Meinecke2022} leading to innovative science and publications in high-impact journals.
For pulsed-power experiments, \emph{FLASH} has been able to reproduce past analytical models \cite{Slutz2001}, is being applied in the modeling of capillary discharge plasmas \cite{Cook2020}, and is being validated against gas-puff experiments at CESZAR \cite{Conti2020PRAB}.
The Flash Center is also collaborating with Los Alamos National Laboratory (LANL) in the modeling of laser-driven experiments of cylinder implosions \cite{Sauppe2020} at the Omega Laser Facility at the University of Rochester and the National Ignition Facility at Lawrence Livermore National Laboratory, in a successful integrated inertial confinement fusion (ICF) verification and validation (V\&V) effort with \emph{xRAGE}. \cite{Gittings2008, Sauppe2023}

The Multi-block Arbitrary Coordinate Hydromagnetic (\emph{MACH2}) code \cite{Frese1987} is a single-fluid, multi-material, three-temperature resistive MHD code, developed by the Center for Plasma Theory and Computation at the Air Force Research Laboratory (AFRL), Phillips Research Site.
It solves the usual set of MHD equations: mass conservation, momentum conservation, electron, ion, and radiation energy, and Faraday's law of induction for the magnetic field.
One fundamental difference between \emph{FLASH} and \emph{MACH2} lies in the formulation of the total energy equation.
Although \emph{MACH2} advances the total energy in a non-conservative manner, this has been proven to not impact the code's ability to capture MHD shocks, provided that adequate grid resolution is used. \cite{Velikovich2012}
Radiation is calculated using a single energy group (Gray radiation), with a flux-limited, non-equilibrium model.
The EOS and transport coefficients (opacities, thermal conductivities, magnetic resistivity) can be obtained from the LANL SESAME tables.
The code also contains options to use a gamma-law EOS and certain analytical transport coefficients (e.g., Spitzer thermal conductivity). 

This code has an adaptive mesh algorithm, which can alter the computational grid every time step according to user-specified criteria.
Its Arbitrary Lagrangian–Eulerian (ALE) framework allows simulations to be run in pure Lagrangian, pure Eulerian, or a combination of the two methods.
In the pure Eulerian mode, the code is still taking a Lagrangian step, but maps the result back to the fixed Eulerian mesh.
The grid spacing is potentially adjusted by the adaptive algorithm, depending on magnetic or fluid pressure gradients (or both), which can provide increased accuracy in regions of interest while saving computational time.
The Eulerian method, where the computational grid is fixed in space for the entire duration of a simulation, is perhaps the easiest to conceptualize and analyze.
However, it may require increased resolution in certain regions to properly model important phenomena driving the system dynamics.
New \emph{MACH2} results in this work use the pure Eulerian method for comparison with \emph{FLASH}.

\emph{MACH2} contains a self-consistent circuit model, which is intended to represent the refurbished Z pulsed power machine at Sandia National Laboratory (SNL). \cite{McBride2010}
The input open-circuit voltage profile and other circuit parameters are described in a previous paper. \cite{Wessel2015}
This same circuit model is also now implemented in \emph{FLASH}.

\emph{MACH2} has been successfully used for a variety of studies, which supports its use as a code that has gone through an extensive amount of V\&V.
These studies include, but are not limited to, explosive magnetic generators, plasma opening switches,~\cite{Buff1987,Degnan1987} compact toroid schemes,~\cite{Peterkin1993,Degnan1993,Peterkin1995prl} ICF and alternative fusion concepts,~\cite{Peterkin1995} and Z-pinches with solid liners. \cite{Degnan1995,Shumlak1995}
Some have questioned whether previous SZP simulations used \emph{MACH2} correctly with appropriate boundary conditions and sufficient spatial resolution.
The code-to-code comparisons reported in this paper are intended to build confidence that these codes can accurately model Z-pinches and help guide experiments.

\emph{MACH2} is also actively used and developed at the Naval Research Laboratory (NRL), and newer versions of the code may have significant differences from the version used in this study.
One of the purposes of this work is to assess the SZP platform with \emph{FLASH} within the context of MIFTI's previous and current research using the version of \emph{MACH2} in their possession.
Therefore, it is not our intention to fix any potential errors that may be discovered in \emph{MACH2}.
Any code-to-code discrepancies described in this work pertain specifically to this version of \emph{MACH2} and should not have any bearing on newer versions of the code being used by other research groups.

\section{\label{sec:tests}Analytical Tests}
In published \emph{MACH2} simulations of the SZP, shock waves were identified as crucial in preheating the target plasma and piling up liner mass at the liner/target interface. \cite{Ruskov2020} 
The interpretation of these shock waves has come under some criticism in recent years. \cite{Lindemuth2018,Lindemuth2021}
Nevertheless, these shock waves are present in the MACH2 simulations and are complex phenomena as they develop in a magnetized medium with important radiative effects.
For these reasons, we decided to test both \emph{FLASH} and \emph{MACH2} with simpler analytical problems in which Z-pinch-relevant shock physics is important.
In the subsections that follow, we present test results from a radiative shock problem and from the cylindrical Noh problem.
One purpose of these tests is to help build confidence in each code's ability to accurately model constituents that make up the fluid- and thermo-dynamics of the complicated SZP.

\begin{figure}[H]
    \centering
    \includegraphics[angle=270,width=0.58\linewidth]{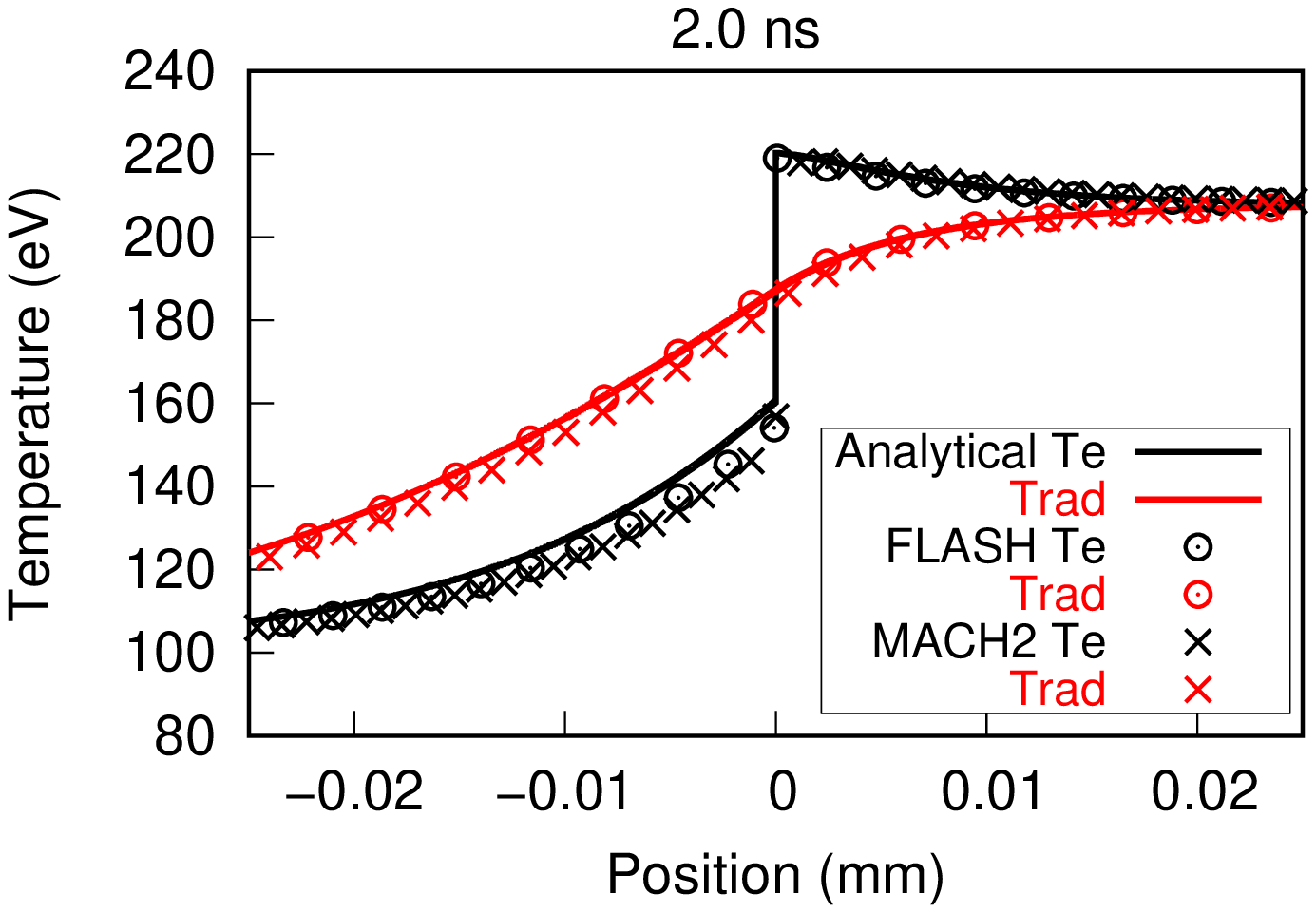}
        \includegraphics[angle=270,width=0.58\linewidth]{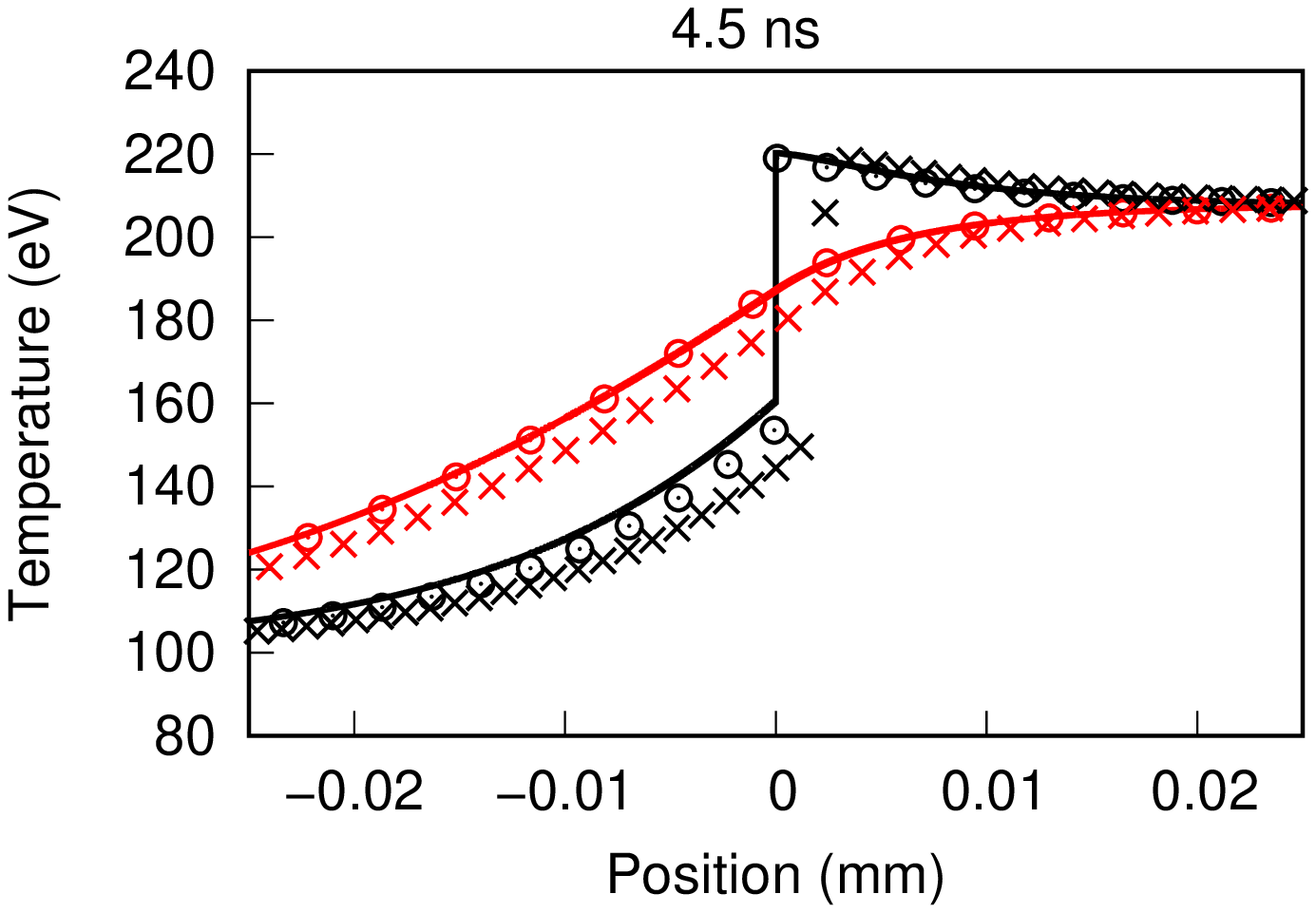}
            \includegraphics[angle=270,width=0.58\linewidth]{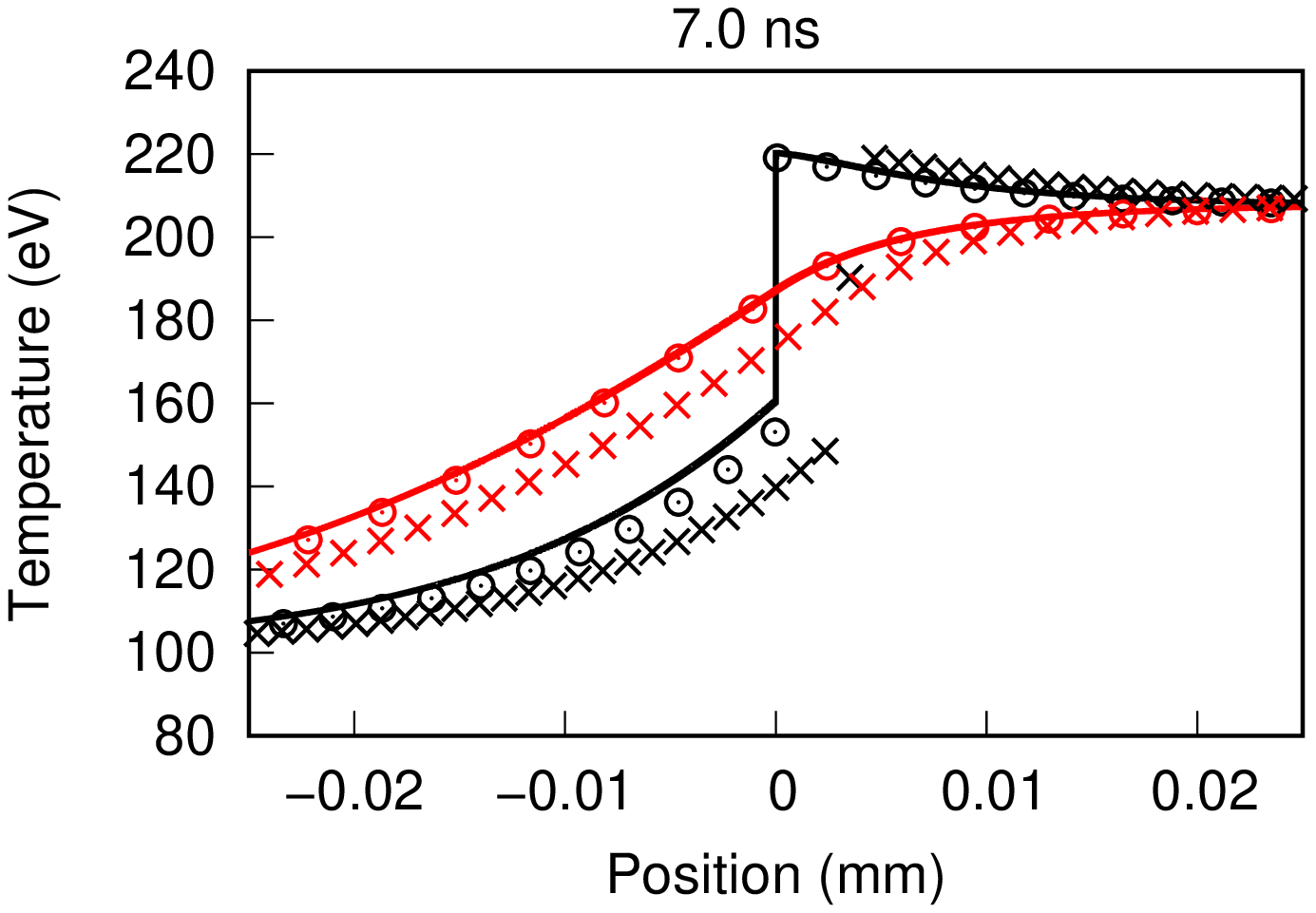}
    \caption{Analytical solution of electron (same as ion, black) and radiation temperatures (red) for the radiative shock test problem as compared to \emph{FLASH} and \emph{MACH2} simulation results.
    The analytical solution is shown as a solid line whereas the \emph{FLASH} and \emph{MACH2} results are circle-dot and cross symbols, respectively. 
    The top panel (2 ns) appears to show good agreement, but for later times (4.5 ns and 7 ns) we see an increasingly discrepant position in the \emph{MACH2} result.}
    \label{fig:RadShock}
\end{figure}

\subsection{\label{subsec:RadShock}Radiative Shock Problem}
Radiative shocks (and radiation in general) are essential elements of SZP simulations. 
As reported by \citet{Ruskov2021b} (cf. Fig.~9 therein), \emph{MACH2} simulations indicate that a more radiative (higher \textit{Z}) liner is compressed more for the same liner mass and driver; consequently coupling its kinetic energy into target internal energy more efficiently; ultimately resulting in higher yield.

The radiative shock problem presented here follows the prescription described in \citet{Lowrie2008}, and its simulation setup is also described in the \emph{FLASH} user's guide. \cite{flashug}
It is an analytical solution to a 1D, steady, radiative shock in which electron and ion temperatures are in equilibrium, but the radiation temperature differs.
Constant opacities are used for a single radiation energy group (gray).
The purpose of the problem is to test a code's radiation transfer and shock-capturing capabilities, both of which are important for  the modeling of the SZP concept.

The Planck opacities (absorption and emission) are set to approximately 423 cm\textsuperscript{-1}, and the Rosseland opacity (transport) is approximately 788 cm\textsuperscript{-1}.
An ideal EOS is used with adiabatic index $\gamma = 5/3$, atomic number $Z = 1$ (also the constant ionization state), and atomic mass $A = 2$~amu.
Electron, ion, and radiation temperatures are initially in equilibrium, and their upstream (pre-shock) value is 100 eV.
The electron and ion temperatures remain in equilibrium throughout the domain for the duration of the simulation due to heat exchange with an enforced reduction to the equilibration time, but radiation temperature can change.
The upstream density is set to 1.0 g~cm\textsuperscript{-3}, and the remaining upstream and all downstream conditions are set appropriately to maintain a steady shock with Mach number $M=2$.

Fig.~\ref{fig:RadShock} shows the analytical solution for electron (same as ion) and radiation temperatures as well as simulation results from both codes.
The grid resolutions used for these simulations were approximately $\mathrm{0.146~\mu m}$ and $\mathrm{0.293~\mu m}$ for \emph{FLASH} and \emph{MACH2}, respectively.
A resolution convergence study was conducted with \emph{MACH2}, and the results did not change with higher resolution.
Note that Fig.~\ref{fig:RadShock} does not contain every data point from either code to avoid over-crowding the plot.
At a relatively early time of 2 ns, it appears that both \emph{FLASH} and \emph{MACH2} recover the analytical solution, providing confidence that the radiation transfer algorithms give accurate results.
However, at later times, we see that only \emph{FLASH} captures the exact position of the shock, whereas \emph{MACH2} shows an increasing positional offset in the electron/ion temperature jump.
This result shows that \emph{MACH2} does not capture electron-radiation coupling as accurately as \emph{FLASH}, especially in the presence of shocks.
This kind of discrepancy will play a role in the SZP1* simulations presented later in this work (Section~\ref{sec:SZP1}).

In this version of \emph{MACH2}, the radiation transport calculation occurs before the hydrodynamic advection calculation, which is a typical operator-split approach.
However, the \emph{MACH2} result shows a small positional error on the order of a fraction of a cell width, which accumulates over time, hence the position of the shock continuously drifts farther away from the analytical solution at later times.
\emph{FLASH} utilizes a similar operator-split approach, but does not exhibit the same error.
Additional tests were conducted to match the order of operations, but the results did not change.
It is also important to note that \emph{MACH2} took on the order of 100,000 computational time steps for the radiative shock test problem, but the SZP1* simulation discussed later, required over 2.6 million steps, thus the aforementioned error accumulation could be significant in the SZP1* simulation.

\subsection{\label{subsec:Noh}Magnetized Noh Problem}
The classical Noh problem \cite{Noh1987} provides a test to benchmark the accuracy of hydrodynamic codes to capture shock dynamics in a convergent geometry.
Initially, an infinite mass is set with a homogeneous inward velocity in cylindrical coordinates.
Due to the singularity at the origin, an accretion shock wave is generated that propagates outwards, decelerating the incoming fluid mass. 
The magnetized extension of the problem, derived by \citet{Velikovich2012}, is well-reproduced by \emph{FLASH}. \cite{Tzeferacos2012}
We have repeated the simulation here since we are using a newer version of \emph{FLASH}, and we compare with results from the relevant version of \emph{MACH2} used in the published SZP simulations \cite{Rahman2009Plasma,Wessel2015,Wessel2016,Ruskov2020}.
A simulation setup for this test is also included in the release version of \emph{FLASH} and described in the user's guide. \cite{flashug}

\begin{figure}[H]
    \centering
    \includegraphics[width=0.6\linewidth]{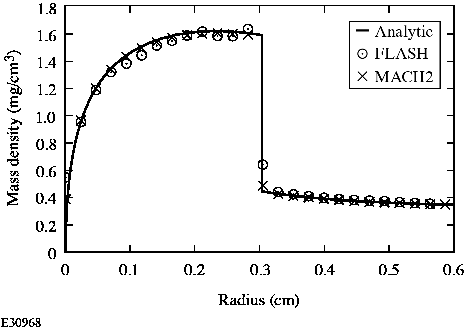}
    \caption{Analytical solution of mass density to the magnetized Noh problem as compared to \emph{FLASH} and \emph{MACH2} simulation results.
    The analytical solution is shown as a solid line whereas the \emph{FLASH} and \emph{MACH2} results are circle-dot and cross symbols, respectively.
    Both codes recover the expected profile.}
    \label{fig:Noh}
\end{figure}

It was observed that, for both codes, the numerical results converge towards the analytical solution with increased resolution.
Fig.~\ref{fig:Noh} shows that both codes approximately reproduce the analytical solution for mass density.
The key features that are modeled accurately are the peak density and the location of the shock.
Note that this test only involves hydromagnetic advection, and therefore, the \emph{MACH2} result does not suffer from the same error accumulation issues observed in the radiative shock test.
The accuracy of this test builds confidence in \emph{FLASH} and \emph{MACH2} simulations of cylindrical MHD implosions, which is clearly relevant to the SZP concept. 

\section{\label{sec:SZP2}Silver Liner Models}
We first present simulations of the silver liner on a DT target configuration proposed in \citet{Wessel2015}, referred to as SZP2.
This configuration was chosen because it was used in \citet{Ruskov2020} in response to SZP criticism,~\cite{Lindemuth2018} and is therefore employed here in the \emph{FLASH} calculations.

Our setup consists of a flat density profile for both fuel  ($\rho_{\mathrm{fuel}}=\mathrm{9.8\times 10^{-3}~g\,cm^{-3}}$) and liner ($\rho_{\mathrm{liner}}=\mathrm{0.6~g\,cm^{-3}}$).
Initially, the fuel region extends from $r = 0$ to $r = 0.2~\mathrm{cm}$, and the thickness of the liner is $0.1~\mathrm{cm}$.
The computational domain consists of a uniform grid of 1,024 points that extends to $r = 0.4~\mathrm{cm}$. 
This ensures a resolution of 10 cells to describe the fuel region at stagnation.
At $t=0$, the system is in thermal equilibrium at an initial temperature of $2~\mathrm{eV}$.
In this first comparison, we have taken an \emph{ideal-EOS} approach, in which a gamma-law EOS with $\gamma = 5/3$, constant ionization of $Z=1$ for the fuel and $Z=10$ for the liner, and analytical formulas for radiation opacities corresponding to free-free electron transitions (Bremsstrahlung radiation) \cite{zel2002physics,Spitzer1962} are assumed.
More precisely, we are taking the Planck mean opacity $K_P^{ff}$ for the emission and absorption opacities and the Rosseland mean opacity $K_R^{ff}$ for the transport opacity, where $K_P^{ff}$ and $K_R^{ff}$ are (in c.g.s.):

\begin{equation}
    \label{eq:Planck}
    K_P^{ff} = 0.43 \dfrac{Z^3}{A^2} \dfrac{\rho ^2_{\text{g\,cm}^{-3}}}{T^{7/2}_{\text{keV}}}\text{ cm}^{-1},
\end{equation}

\begin{equation}
    \label{eq:Rosseland}
    K_R^{ff} = 0.014 \dfrac{Z^3}{A^2} \dfrac{\rho ^2_{\text{g/cc}}}{T^{7/2}_{\text{keV}}} \text{ cm}^{-1}.
\end{equation}

\noindent Here, $A$ refers to the mass number. 

The motivation behind the \emph{ideal-EOS} approach is to verify that \emph{FLASH} can accurately solve for the fundamental physics that model and govern a pinch implosion.
Particularly, appropriate treatment of the vacuum region is essential in an Eulerian code like \emph{FLASH}.
The vacuum region is modeled as a low-density fluid whose task is to transfer the magnetic field from the outer boundary, placed at $r=0.4~\mathrm{cm}$, to the outer surface of the liner while adhering to a current-free profile, $B \sim 1/r$.
To ensure this, an artificially high value of magnetic diffusivity in the vacuum region was used, namely, $\eta _\mathrm{vac}\sim 10^{11} \mathrm{~cm^2\,s^{-1}}$.
Additionally, a temperature ceiling was imposed in the vacuum to avoid potential build-up of thermal pressure that could affect pinch dynamics.
The temperature ceiling has the side benefit of keeping thermal conduction low in the vacuum, hence reducing liner-vacuum heat losses.

\begin{figure}[H]
    \centering
    \includegraphics[width=0.6\linewidth]{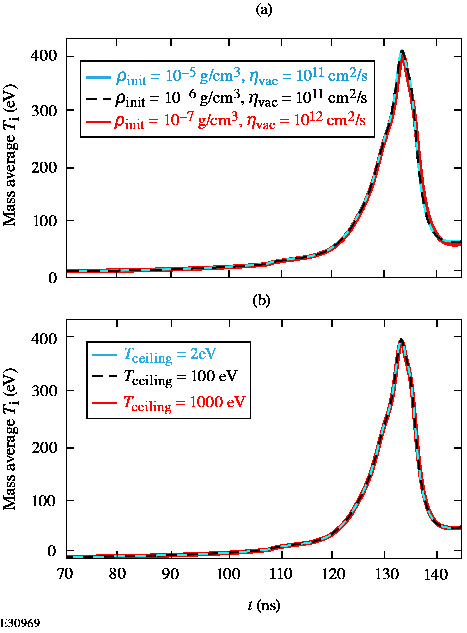}
    \caption{Mass-averaged fuel ion temperature vs. time in the \emph{ideal-EOS} SZP2 run.
    (a) Different initial densities in the vacuum, spanning $10^{-7}-10^{-5}$ g  cm$^{-3}$, and vacuum diffusivity $\eta _{\mathrm{vac}}$ ($10^{11}-10^{12}$ cm$^2$~s$^{-1}$.
    (b) Different temperature ceilings in the vacuum, ranging from 2 eV to 1,000 eV.}
    \label{fig:SZP2TivsTvac}
\end{figure}

A signature of correct behavior of the vacuum is the robustness of the implosion dynamics to changes in the parameters that model the vacuum region.
This is shown in Fig.~\ref{fig:SZP2TivsTvac}, where the temporal evolution of mass-averaged ion temperature is minimally affected by the value of the initial vacuum density (provided that it is sufficiently low), vacuum diffusivity (provided that it is sufficiently high), and temperature ceiling.

\begin{figure}[H]
    \centering
    \includegraphics[width=0.7\textwidth]{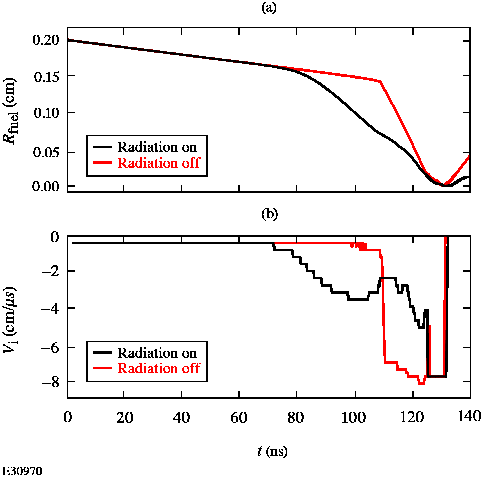}
    \caption{Fuel radius (a) and implosion velocity $V_i$ (b) as a function of time in the \emph{ideal-EOS} SZP2 run.
    The black line denotes a simulation in which the radiation transport operates normally, whereas the red line denotes the simulation in which the radiation transport is artificially switched off.}
    \label{fig:SZP2LinerTrajectory}
\end{figure}

\begin{figure}[H]
    \centering
    \includegraphics[width=0.7\textwidth]{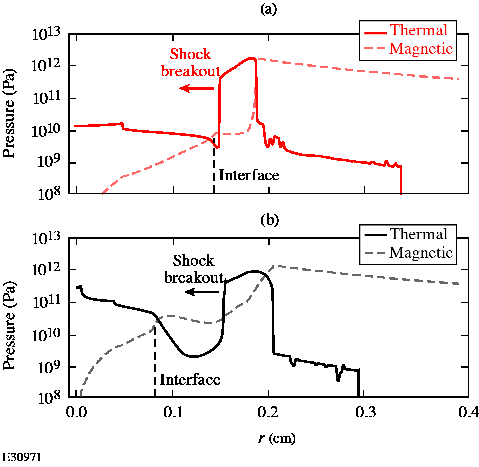}
    \caption{Profiles of thermal pressure (solid) and magnetic pressure (dashed) at the time of shock breakout in the \emph{ideal-EOS} SZP2 runs, for (a) radiation physics turned off ($t = 109 \mathrm{~ns}$), and (b) radiation physics turned on ($t = 107 \mathrm{~ns}$).}
    \label{fig:SZP2Profiles}
\end{figure}

The dynamics of the implosion are sketched in Fig.~\ref{fig:SZP2LinerTrajectory}, where the evolution of the fuel radius and the implosion velocity are shown for runs with radiation physics on (black) and off (red).
Initially, the fuel is slowly compressed as a result of a pressure imbalance, present in the initial conditions. 
This is a result of the three regions (fuel, liner, and vacuum) being initialized with a homogeneous temperature of 2 eV. 
At $t = 80  \mathrm{~ns}$, the trajectories of the two runs depart.  
In the run with radiation physics turned off, a jump-off velocity of the liner of $7 \mathrm{~cm/\mu s}$ is observed when the leading shock breaks out into the fuel at $t = 109$ ns.
This is consistent, albeit slightly above, the $\sim 6 \mathrm{~cm/\mu s}$ implosion velocity reported both in \citet{Lindemuth2018}, cf. Fig.~4, and in \citet{Ruskov2020}, cf. Fig.~3(b). 
Thermal and magnetic pressure profiles at the time of shock breakout are depicted in Fig.~\ref{fig:SZP2Profiles}(a) for this run.

The run where radiation physics is turned on shows more complex dynamics.
Radiation keeps the liner colder, which allows for more magnetic field to diffuse into the fuel.
At its interface with the liner, significant magnetic pressure is built up, and the magnetic piston thereby formed drives the initial stages of fuel compression.
This stage occurs between $t = 80 \mathrm{~ns}$ and $t = 122 \mathrm{~ns}$.
In Fig.~\ref{fig:SZP2Profiles}(b), the thermal and magnetic pressure profiles are shown at an intermediate time of $t = 107 \mathrm{~ns}$, which corresponds to the moment of shock breakout (i.e., when the shock traverses the fuel/liner interface).
After this time, the liner travels inward faster than the magnetic piston driving the compression.
Subsequently, it catches up with the fuel at $t = 122 \mathrm{~ns}$ and sets the implosion velocity at a higher value, similar to the radiation-off case.

After shock breakout, the shock travels in the fuel, heating it non-adiabatically (viz., shock preheating).
Eventually, the shock reaches the symmetry axis, rebounds off, and propagates outward in the form of a weaker shock or sound wave.
Subsequent shock preheating can take place due to shock rebound at the inner surface of the liner, or due to additional shocks launched by the driver.
Shock preheating ends when the fuel temperature is raised to a value where further compression becomes subsonic. 

\begin{figure}[H]
    \centering
    \includegraphics[width=0.7\linewidth]{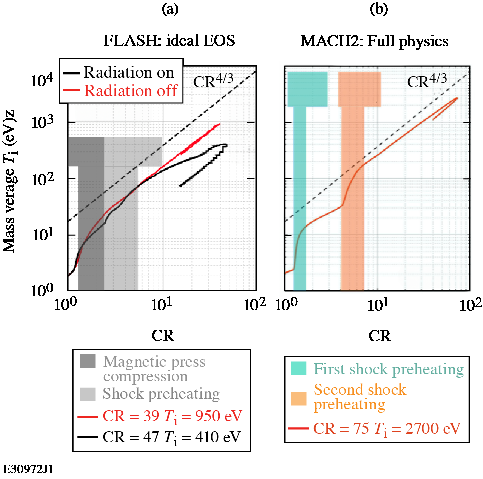}
    \caption{Mass-averaged ion temperature vs. convergence ratio (CR) in the \emph{ideal-EOS} \emph{FLASH} SZP2 run (left) and \emph{full-physics} \emph{MACH2} run (right) from \citet{Ruskov2020}, where SESAME tables were used.}
    \label{fig:SZP2TivsCR}
\end{figure}

The evolution of the ion temperature as a function of the convergence ratio provides insights into the implosion dynamics.
Throughout this work, the convergence ratio (CR) is defined as the ratio of the initial outer fuel radius to the compressed outer fuel radius.
This is plotted in Fig.~\ref{fig:SZP2TivsCR} and compared to its counterpart \emph{full-physics} \emph{MACH2} run from \citet{Ruskov2020}.
Highlighted in light gray is the stage of the implosion driven by magnetic pressure, whereas the darker gray region denoted the period during which the main shock propagates in the fuel.
It can be seen that, despite the different early dynamics previously described, both radiation-off and radiation-on simulations depart from approximately the same fuel temperature after shock preheating, $CR = 5$.
At the later stage of subsonic compression, the the radiation-off run closely follows an adiabatic trajectory, while the radiation-on run deviates significantly.
This indicates that radiation losses dominate over thermal conduction losses in the SZP2 configuration, as was anticipated in \citet{Lindemuth2018}, cf. Table III.

Comparing the \emph{MACH2} and \emph{FLASH} simulations, it can be observed that the fuel adiabat after shock preheating is significantly lower in \emph{FLASH}, resulting in lower fuel temperatures and a lower convergence ratio at stagnation.
In this aspect, the \emph{FLASH} run is more consistent with the results reported in \citet{Lindemuth2018}, cf. Fig.~7: stagnation temperatures slightly below 1 keV at $CR \sim 50$.
However, it should be highlighted that adopting an \emph{ideal-EOS} framework resulted in different early dynamics.
In the \emph{MACH2} run, the occurrence of secondary shock preheating, absent in the simulations in \citet{Lindemuth2018}, allows the fuel adiabat to rise and attain fusion conditions.
In the \emph{FLASH} run, only primary shock preheating is observed, preceded by compression due to magnetic pressure.

\section{\label{sec:SZP1}Xenon Gas-puff Liner Models}
We have developed a new configuration of a xenon gas-puff liner to enact a direct comparison between \emph{FLASH} and \emph{MACH2} simulations of the SZP platform.
Previously published SZP models of Xe gas-puff liners, sometimes referred to as SZP1 \cite{Rahman2009Plasma}, use different initial conditions. 
To avoid confusion, we refer to this new configuration as SZP1*.
The main difference is a lower liner density, which was chosen to generate a faster implosion and thus stronger shocks in the hopes of achieving higher temperatures.

Another motivating factor for SZP1* is that the dominant thermal loss mechanism is thermal conduction rather than radiation as in SZP1.
We can estimate the ratio of thermal conduction losses to radiation losses for DT with the following formula:

\begin{equation}
    \label{eq:losses}
    \frac{Q_{cond}}{Q_{rad}} \approx T_e^3 \bigg(\frac{CR}{50}\bigg)^2 K(\chi_e) ~ n_i^{-2},
\end{equation}

\noindent where $T_e$ is electron temperature in keV, $CR$ is fuel convergence ratio, $K$ is a coefficient that accounts for the effect of the electron Hall parameter, $\chi_e$, on the electron thermal conductivity, and $n_i$ is the ion number density in 10\textsuperscript{24} cm\textsuperscript{-3}.
This equation is derived from the work of \citet{Lindemuth2009} and represents an estimate of the ratio of the rate of electron thermal conduction to the rate of Bremsstrahlung radiation, specifically for DT.
For hydrogen, $K(\chi_e) = (4.664\chi_e^2 + 11.92) / (\chi_e^4 + 14.79 \chi_e^2 + 3.77)$ was used.
The input parameters (from the \emph{FLASH} simulations) and resulting ratios for SZP1 and SZP1* are summarized in Table~\ref{tab:losses}.
For the original SZP1 scheme, we calculated this ratio to be $< 1$ near stagnation, making radiation the primary heat loss mechanism.
Conversely, for SZP1*, we estimate this ratio to be $> 57$, thus thermal conduction losses dominate near stagnation.
This configuration is potentially advantageous because thermal conduction losses can be reduced if sufficient magnetic field is diffused into the fuel.
The main contributing factors for this difference, according to Eq.~\eqref{eq:losses} and Table~\ref{tab:losses}, are the $CR$ and $n_i$ attained.
Since SZP1* reaches higher $CR$ than SZP1 because of its lower density, the different thermal loss regimes are ultimately a result of the different densities.

\begin{table}[H]
\centering
\begin{ruledtabular}
\begin{tabular}{cccccc}
Model & $T_e$ (keV) & $CR$ & $\chi_e$ & $n_i$ (10\textsuperscript{24} cm\textsuperscript{-3}) & $Q_{cond}/Q_{rad}$ \\
\hline
SZP1 & 1.09 & 127.7 & 0.23 & 12.4 & 0.147 \\
\hline
SZP1* & 1.93 & 387.7 & 0.19 & 4.57 & 57.1 \\
\end{tabular}
\end{ruledtabular}
\caption{Input parameters from \emph{FLASH} simulations of SZP1 and SZP1* for Eq.~\eqref{eq:losses} and the resulting ratios.}
\label{tab:losses}
\end{table}

\begin{figure}[H]
    \centering
    \includegraphics[width=0.6\linewidth]{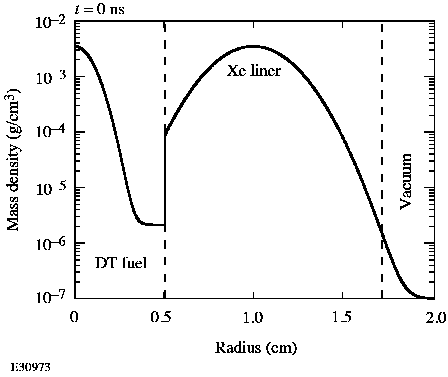}
    \caption{Initial mass density of the new SZP1* configuration.}
    \label{fig:SZP1rho}
\end{figure}

The SZP1* models shown in this section are initialized with the following density profile:
\begin{equation}
    \rho(r) = (\rho_{max} - \rho_{min}) \exp{(-k_r (r - r_0)^2)} + \rho_{min}, \label{SZP1rho} 
\end{equation}
with $\rho_{max} = \mathrm{3.5 \times 10^{-3}~g\,cm^{-3}}$, $\rho_{min} = \mathrm{2.132 \times 10^{-6}~g\,cm^{-3}}$, $k_r = \mathrm{70~cm^{-2}}$, and $r_0 = 0$ for the DT fuel from $r = 0$ to $r = 0.504~\mathrm{cm}$.
The Xe liner region uses $\rho_{max} = \mathrm{3.5 \times 10^{-3}~g\,cm^{-3}}$, $\rho_{min} = \mathrm{1 \times 10^{-7}~g\,cm^{-3}}$, $k_r = \mathrm{15~cm^{-2}}$, and $r_0 = 1~\mathrm{cm}$ from $r = 0.504~\mathrm{cm}$ to $r = 1.722~\mathrm{cm}$.
The tail of the liner Gaussian from $r = 1.722$ to $r = 2.016~\mathrm{cm}$ is modeled as a ``vacuum'' region, with a floor density of $\rho_{min} = \mathrm{1 \times 10^{-7}~g/cm^{-3}}$ enforced for the duration of the simulation.
Fig.~\ref{fig:SZP1rho} shows the initial mass density profile for the SZP1* configuration.

Electron and ion temperatures are initialized at 2 eV everywhere, and the vacuum is forced to remain at this temperature. 
The vacuum is not allowed to emit or absorb radiation, so radiation simply passes through it and leaves the domain.
Electron and ion thermal conductivity is set to zero in the vacuum and the magnetic diffusivity is  constant at $\mathrm{1 \times 10^{12} cm^2/s}$, while the DT fuel and Xe liner regions use magnetic field-dependent Spitzer values \cite{Spitzer1962} for electron and ion thermal conductivity and magnetic resistivity.
Both \emph{MACH2} and \emph{FLASH} choose which transport coefficient model to use (constant or Spitzer) based on material mass fractions, which become important at material interfaces where mixed cells can exist.
\emph{MACH2} uses a front-tracking algorithm that leads to interfaces with only one mixed cell, whereas \emph{FLASH} does not yet have this capability, so the material interfaces are more diffused and typically have 3-4 mixed cells.
In \emph{FLASH}, we have chosen to treat a mixed cell as vacuum when its vacuum mass fraction is greater than or equal to 99\%, which maintains a sharp one-cell vacuum/liner interface at this transition.
It is also important to note that tests with \emph{MACH2} were conducted without front-tracking, and there was not a qualitative difference in behavior or results.
The EOS, ionization state, and opacities are taken from \emph{PROPACEOS} tables for DT and Xe.

In \emph{FLASH}, the domain is a fixed grid of 20,160 computational cells, resulting in a resolution of $\delta r = \mathrm{1~\mu m}$.
Due to computational limitations, the \emph{MACH2} model used fewer computational cells, but we opted for non-uniform cell widths to maintain an adequate number of cells near the symmetry axis as the implosion reaches stagnation ($\Delta r = 1~\mathrm{\mu m}$ near $r=0$).
The lower-$r$ boundary condition is axisymmetric, and the upper-$r$ boundary condition is open with an enforced density of $\mathrm{1 \times 10^{-7}~g/cm^{-3}}$.
At the upper-$r$ boundary we also use Ampere's Law to set the azimuthal magnetic field whose current is calculated from a circuit model approximating the Z-Machine at Sandia National Laboratories. \cite{McBride2010}

\subsection{\label{subsec:SZP1a}Comparison with \emph{MACH2}}
The \emph{MACH2} code does not have a multi-group radiation diffusion capability, thus, for the purposes of this comparison, we ran the SZP1* setup with a single radiation group spanning photon energies from 0.1 eV to 10\textsuperscript{5} eV.
With the use of a table conversion tool, both codes were able to use the same tabulated data (single-group \emph{PROPACEOS} tables for DT and Xe) for EOS, ionization state, and opacities.

\begin{figure}[H]
    \centering
    \includegraphics[width=0.9\linewidth]{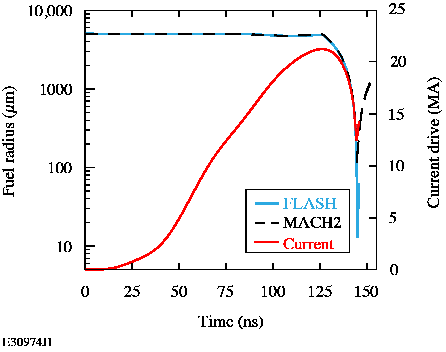}
    \caption{Comparison of shell trajectories (i.e., fuel outer radius) from SZP1* simulations with \emph{MACH2} (dashed black) and \emph{FLASH} (solid cyan).
    The load current resulting from the circuit model in \emph{FLASH} (red) is also shown.}
    \label{fig:SZP1traj}
\end{figure}

Fig.~\ref{fig:SZP1traj} shows the trajectory of the outer fuel radius from runs with both codes as well as the load current (for reference) from the circuit model in the \emph{FLASH} simulation.
The magnetic field builds up for several ns to increase enough to push the inner surface of the liner and start compressing the fuel. This is followed by relatively short sequences of rapid acceleration, stagnation, and expansion.
In the \emph{FLASH} simulation, the load current peaks at about 21.27 MA near 126 ns, which approximately corresponds to the beginning of the acceleration phase and $\sim 19$ ns before stagnation.
Although the timing of implosion dynamics matches fairly well, the \emph{FLASH} model converges significantly more than the \emph{MACH2} model.
The convergence ratio reached in the \emph{FLASH} simulation ($CR = 388$) is more than four times higher than that of the \emph{MACH2} result ($CR = 87$).

\begin{figure}[H]
    \centering
    \includegraphics[width=0.6\linewidth]{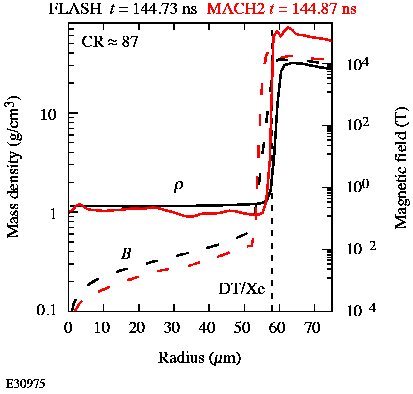}
    \caption{Comparison of mass density (solid) and magnetic field (dashed) at $CR=87$ from SZP1* simulations from \emph{MACH2} (red) and \emph{FLASH} (black).
    The fuel/liner interface is marked by a short-dash vertical line.}
    \label{fig:SZP1peakBphi}
\end{figure}

The mass density and magnetic field profiles at $CR=87$ are shown in Fig.~\ref{fig:SZP1peakBphi}.
This is the maximum $CR$ attained in the \emph{MACH2} run, whereas the \emph{FLASH} simulation continues to compress.
While a cursory inspection of Fig.~\ref{fig:SZP1peakBphi} may conclude that the simulations match fairly well, there are a two key differences to note.
The liner in the \emph{MACH2} simulation has compressed to larger densities than in the  \emph{FLASH} simulation. Further, in the \emph{MACH2} simulation we see a significant built-up of magnetic field just inside the fuel abutting the fuel/liner interface, which lowers thermal conductivity, insulates the fuel, and reduces thermal losses. This disparity in magnetic field accumulation in the fuel is identified as the main cause for the observed difference in the maximum convergence ratios, at stagnation, between simulations.

\begin{figure}[H]
    \centering
    \includegraphics[width=0.6\linewidth]{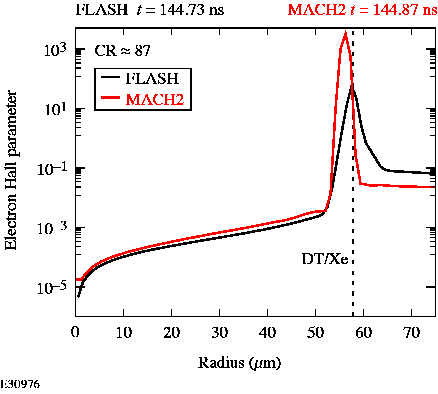}
    \caption{Comparison of the electron Hall parameter at $CR=87$ from SZP1* simulations with \emph{MACH2} (red) and \emph{FLASH} (black).
    The fuel/liner interface is marked by a short-dash vertical line.
    The fuel adjacent to the fuel/liner interface in the \emph{MACH2} simulation is significantly more magnetized than in the \emph{FLASH} result.}
    \label{fig:SZP1peakChi}
\end{figure}

The relatively larger magnetic field in \emph{MACH2} leaks into the fuel at approximately the same time or shortly after the main shock in the liner breaks out into the fuel.
This occurs relatively early ($\sim127$ ns) at $CR \sim 1.11$, and the simulations begin to diverge after this point.
After shock breakout, the \emph{MACH2} simulation predicts a thin, cold region in the fuel next to the liner. The temperature drop leads to an increase in magnetic resistivity which, in turn, allows more magnetic field to diffuse inwards, further inhibiting thermal conduction.
In the limit of large magnetization, the perpendicular thermal conductivity is proportional to $T_e^{2.5} / \chi_e^2$, where $\chi_e$ is the electron Hall parameter.
This thin fuel region next to the liner is more magnetized in the \emph{MACH2} simulation than in the \emph{FLASH} run as shown in Fig.~\ref{fig:SZP1peakChi}, with peak values of $\chi_e \approx 3138$ and $\chi_e \approx 58.55$, respectively.
Taking also into account the different temperatures, we estimate that the thermal conductivity in this part of the fuel is more than 120 times greater in the \emph{FLASH} simulation than in the \emph{MACH2} run.
This observation explains why thermal conduction losses are higher in the \emph{FLASH} simulation and is consistent with the continued compression of the fuel to higher $CR$.
Also note that the magnetic field spike inside the fuel in the \emph{MACH2} result (see Fig.~\ref{fig:SZP1peakBphi}) would require a return current at this location, and we do not generally expect to see return currents inside the fuel in Z-pinches.
Nevertheless, the \emph{MACH2} result shows how increased fuel magnetization can benefit the SZP1* configuration by reducing thermal losses, in turn leading to higher temperatures and larger, more stable $CR$ values.

\begin{figure}[H]
    \centering
    \includegraphics[width=0.6\linewidth]{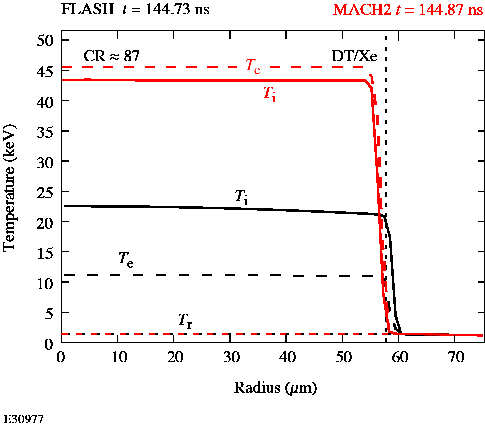}
    \caption{Comparison of electron temperature (dashed), ion temperature (solid), and radiation temperature (dotted) at $CR=87$ from SZP1* simulations with \emph{MACH2} (red) and \emph{FLASH} (black).
    The fuel/liner interface is marked by a short-dash vertical line.
    In the \emph{FLASH} result, the fuel has $T_e < T_i$, whereas in \emph{MACH2}, at stagnation, $T_e > T_i$.
    Also, the temperatures in the \emph{MACH2} run decrease to liner values \emph{before} the interface is reached, while in the \emph{FLASH} profile the temperatures decrease after the interface inside the liner.
    As a result, in the \emph{MACH2} run, the magnetic resistivity and the magnetization of the fuel adjacent to the interface are larger than in the the \emph{FLASH} simulation, insulating the fuel from heat conduction losses.  
}
    \label{fig:SZP1peakTemp}
\end{figure}

The electron, ion, and radiation temperature profiles at $CR=87$ are shown in Fig.~\ref{fig:SZP1peakTemp}.
Here we see a much clearer discrepancy between the two simulations.
The fuel in the \emph{MACH2} run has a much higher electron temperature, which helps explain why the implosion stagnates earlier than in the \emph{FLASH} simulation.
It is also noteworthy that, in the \emph{FLASH} result, we have a fuel whose $T_e < T_i$.
Conversely, in the \emph{MACH2} simulation, at stagnation, $T_e > T_i$.
Generally, in Z-pinch experiments, one may expect the ion temperature to be higher than the electron temperature, since electrons lose energy via radiation, thermal conduction, and heat exchange with the ions, whereas ions are also subject to shock heating.
Nevertheless, the temperature inversion observed in the \emph{MACH2} result is not necessarily nonphysical, given the large fuel magnetization.
In such regimes, ions can be more thermally conductive than electrons, so it is possible for ions to lose more thermal energy and remain colder than electrons.
Also, we again observe a discrepancy at the fuel/liner interface where the temperatures in the \emph{MACH2} run sharply decrease to liner values \emph{before} the interface is reached, while in the \emph{FLASH} profile the temperatures decrease after the interface an inside the liner.
This helps further explain the aforementioned presence of a larger magnetic field values in the fuel in the \emph{MACH2} simulation:
The lower-temperature region just inside the fuel/liner interface results in higher magnetic resistivity, which in turn allows for more magnetic field to diffuse into the fuel.
 
\begin{figure}[H]
    \centering
    \includegraphics[width=0.6\linewidth]{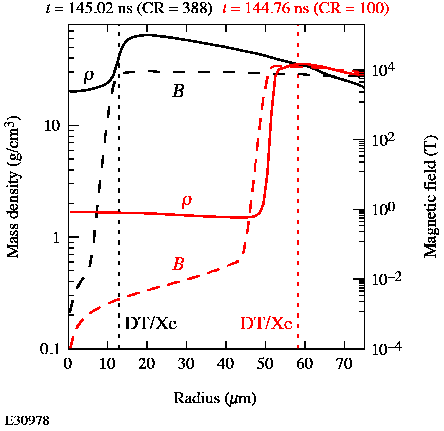}
    \caption{Comparison of mass density (solid) and magnetic field (dashed) from the SZP1* \emph{FLASH} simulation at $CR=100$ (red) and $CR=388$ (black).
    The fuel/liner interfaces are marked by short-dash vertical lines.}
    \label{fig:SZP1stagBphi}
\end{figure}

The \emph{FLASH} model continues to compress and reaches a peak $T_{ion}$ of about 18 keV, on-axis,  at $CR=100$, which occurs at 144.75 ns.
After this peak, the \emph{FLASH} model compresses further, for $\sim 265$ ps, and reaches $CR$ of approximately 388. This latter compression is accompanied by thermal losses that result in lower-than-peak temperatures.
Fig.~\ref{fig:SZP1stagBphi} shows a comparison of the mass density and azimuthal magnetic field from the \emph{FLASH} model at $CR=100$ and $CR=388$ (stagnation).
The fuel density has increased by an order of magnitude, which is consistent with the decrease in volume from a radius of 50 $\mathrm{\mu m}$ to 13 $\mathrm{\mu m}$.
The magnetic field in the fuel has also increased, but the plasma beta is still much larger than unity due to the high thermal pressure.

\begin{figure}[H]
    \centering
    \includegraphics[width=0.6\linewidth]{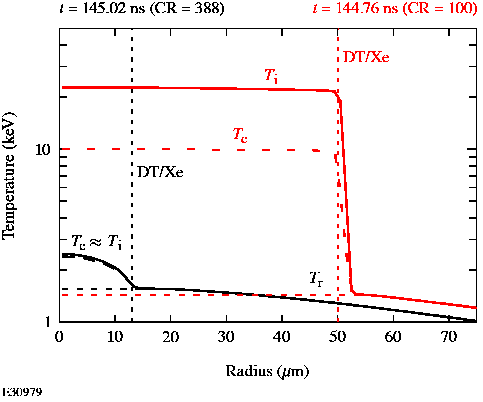}
    \caption{Comparison of electron temperature (dashed), ion temperature (solid), and radiation temperature (dotted) from the SZP1* \emph{FLASH} simulation at $CR=100$ (red) and $CR=388$ (black).
    The fuel/liner interfaces are marked by short-dash vertical lines.}
    \label{fig:SZP1stagTemp}
\end{figure}

Fig.~\ref{fig:SZP1stagTemp} shows a comparison of the electron, ion, and radiation temperatures from the \emph{FLASH} model at $CR=100$ and $CR=388$ (stagnation).
From this comparison, we observe that thermal losses have begun to dominate beyond $CR=100$.
These are primarily due thermal conduction, as we estimated in the previous analysis at the beginning of this section (see Eq.~\eqref{eq:losses} and Table~\ref{tab:losses}).
Meanwhile, density increases due to compression,  eventually causing the fuel to stagnate when the pressure is sufficiently high.
Note that the radiation temperature remains unchanged during this time, and the electron and ion temperatures equilibrate as the fuel stagnates. The temperature inversion seen in the \emph{MACH2} simulations is absent in the \emph{FLASH} run.

Despite having the same initial conditions, circuit model, transport coefficients, and EOS and opacity tables, we were not able to reproduce the \emph{MACH2} result with \emph{FLASH} simulations.
We see that the fuel stagnates at lower $CR$ in the \emph{MACH2} run because the latter reaches much higher temperatures and thus has enough thermal pressure to halt the implosion.
The ability of the fuel to retain its thermal energy (i.e., high temperatures) depends on its thermal losses via radiation and thermal conduction.
The fact that discrepancies start becoming apparent after the shock breakout may call into question the codes' shock-capturing capabilities.
We have shown that the version of \emph{MACH2} used in this work does not reproduce the analytical solution of the radiative shock test problem as accurately as \emph{FLASH} (see Section~\ref{subsec:RadShock}).
The integrated SZP1* simulations are more complicated than the simple benchmark problem, and the ability to accurately model radiative shocks at material interfaces in SZP1* is crucial for accurately predicting thermal conduction losses in the fuel.
The \emph{MACH2} SZP1* simulation should have a similar error accumulation as observed in Fig.~\ref{fig:RadShock}, once the shock breaks out into the fuel.
However the error is potentially larger due to the greater number of computational time steps ($\sim$ 2.65 M).

Another interesting observation from the \emph{MACH2} result is that the fuel electron temperature remains higher than than the ion temperature (see Fig.~\ref{fig:SZP1peakTemp}).
We see the opposite relation in the \emph{FLASH} model, because the electrons are losing more energy without the thin, highly magnetized layer to insulate them.
This layer, observed only in the \emph{MACH2} result, develops after shock breakout, and is therefore susceptible to the errors associated with radiative shock modeling discussed in Section~\ref{subsec:RadShock}.
One would expect the electrons to be radiating, losing energy via thermal conduction, and transferring energy to the ions, while the ions are also subject to compressional heating.
Due to aforementioned thermal loss mechanisms, the \emph{FLASH} model is allowed to reach higher $CR$ values, where thermal conduction losses become even more important.

The \emph{MACH2} code has been successfully used for and validated against several plasma, inertial confinement fusion, and high energy density physics experiments.
However, the modeling of the SZP1* platform, with specific settings to compare with \emph{FLASH}, is a challenging problem for the particular version of \emph{MACH2} used in this work, due to its issues modeling radiative shocks.
This deficiency, in this version of \emph{MACH2}, leads us to conclude that \emph{FLASH} gives more physically sensible results for SZP1*, even though the \emph{FLASH}-predicted $CR$ values are too large to be experimentally stable.

\subsection{\label{subsec:SZP1b}High-fidelity \emph{FLASH} simulations of the SZP1*}
We ran two additional SZP1* models with \emph{FLASH} to determine effects of using higher-fidelity physics implemented in \emph{FLASH}.
These include newer, higher-fidelity transport coefficients \cite{JiHeld2013,Davies2021}, and multi-group radiation diffusion, neither of which are available in \emph{MACH2}.
These newer transport coefficients are more complicated functions of atomic number and the Hall parameter, and they more accurate than Spitzer coefficients.
The multi-group radiation diffusion model also used the newer transport coefficients, as well as 40 radiation energy groups, spanning the same energy range as the single-group (gray) models.
We denote the \emph{FLASH} runs in this subsection as follows: SP is the single-group run with Spitzer transport coefficients (the same run discussed in the previous subsection), DW 1G is the run with the newer transport coefficients and one radiation group, and DW 40G is the run with the newer transport coefficients and 40 radiation groups.

Table~\ref{tab:results} gives a summary of key results in terms of $CR$, stagnation time, and mass-averaged fuel ion temperature at stagnation.
Note that for all \emph{FLASH} simulations, these stagnation temperatures are lower than the peak ion temperatures.
We observe that with the newer coefficients, SZP1* converges slightly faster and to a smaller radius, but the ion temperature is slightly lower.
The multi-group model converges the fastest and to the highest $CR$ values encountered in this work,  $CR \sim 560$.
At stagnation, the multi-group radiation diffusion run is hotter than both single-group \emph{FLASH} runs, but its peak temperature, which occurs prior to stagnation, is lower.

\begin{table}[H]
\centering
\begin{ruledtabular}
\begin{tabular}{cccc}
Model & CR & Stagnation Time (ns) & $\langle T_{ion} \rangle$~(keV) \\
\hline
\emph{MACH2} & 87.0 & 144.87 & 38.96 \\
\hline
\emph{FLASH} single-group (SP) & 388 & 145.02 & 1.967 \\
\hline
\emph{FLASH} single-group (DW 1G) & 504 & 144.96 & 1.934 \\
\hline
\emph{FLASH} multi-group (DW 40G) & 560 & 144.86 & 2.439 \\
\end{tabular}
\end{ruledtabular}
\caption{SZP1* stagnation results from the \emph{MACH2} simulation and three different \emph{FLASH} runs of increasing physics-fidelity.}
\label{tab:results}
\end{table}

\begin{figure}[H]
    \centering
    \includegraphics[width=0.6\linewidth]{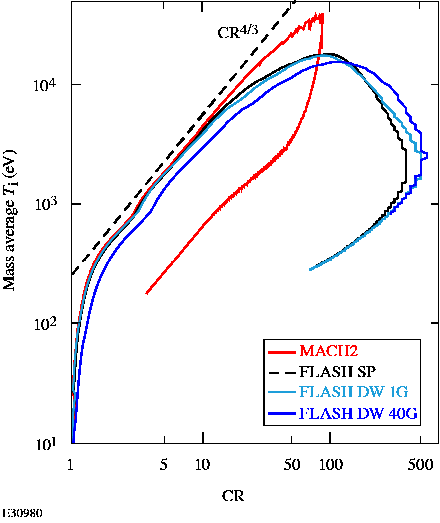}
    \caption{Mass-averaged ion temperature vs. convergence ratio (CR) for all SZP1* simulations from \emph{FLASH} and \emph{MACH2}.
    The single-group \emph{FLASH} run with Spitzer transport coefficients (SP) is directly comparable to the \emph{MACH2} run.
    The other \emph{FLASH} runs used the newer transport coefficients (DW) and used either a single group (1G) or 40 groups (40G).}
    \label{fig:SZP1TionvsCR}
\end{figure}

Fig.~\ref{fig:SZP1TionvsCR} shows the mass-averaged fuel ion temperature as a function of $CR$ for all SZP1* models.
There are several important features to note in this figure:
(1) all models show fuel preheating early ($CR < 2$),
(2) all \emph{FLASH} models continue to compress to higher $CR$ values after peak $\langle T_{ion} \rangle$, whereas the \emph{MACH2} model does not,
(3) the \emph{FLASH} models with newer transport coefficients reach higher $CR$ values, and
(4) the multi-group model is on a lower adiabat and has a lower peak $\langle T_{ion} \rangle$ than all single-group models.
The significance of shock preheating was discussed in Section~\ref{sec:SZP2} in reference to \emph{ideal-EOS} SZP2 models, and similar points apply to SZP1* as well.
However, in SZP1* there is also a radiation wave that provides significant additional fuel preheating.
This was seen when analyzing the early-time behavior of the simulations, and by executing a separate test run with radiation transport switched off, in which the wave was absent. This radiation wave and the initial shock break-out into the fuel effectively set the adiabat of the compression.

Point (2) is essential for understanding the differences between the \emph{FLASH} and \emph{MACH2} models. Thermal losses, which are more significant in the \emph{FLASH} simulations, cool the fuel and allow for higher $CR$ values.
The reasons for the discrepant thermal losses were discussed in the previous subsection.

Points (3) and (4) are specific to the \emph{FLASH} models. 
Use of the newer transport coefficients leads to more thermal conduction losses, which results in higher $CR$ values and lower $\langle T_{ion} \rangle$.
The multi-group model converges slightly more than its single-group counterpart, while its stagnation temperature is higher.
This result indicates that the liner is radiating more efficiently; a colder liner is easier to compress and will subsequently act as a more effective piston for compressing the fuel.
Also, some of the increased liner radiation goes into the fuel, keeping it hot for a longer period.
This speaks to the benefit of using a high atomic-number liner and broadly supports the viability of the SZP concept.

\section{\label{sec:conc}Conclusions}
We have shown that even though \emph{FLASH} and \emph{MACH2} can accurately reproduce analytical benchmarks (with some caveats in the radiative shock problem for \emph{MACH2}), they may not always agree on integrated SZP simulations.
The \emph{ideal-EOS} SZP2 runs from \emph{FLASH} are more consistent with the simulation results of \citet{Lindemuth2018}, where there is no secondary shock preheating and the peak ion temperatures achieved are slightly less than 1 keV.
However, it is important to keep in mind that the \emph{ideal-EOS} framework used for the \emph{FLASH} SZP2 model led to different early-time dynamics.

The SZP1* simulations are a more complete code-to-code comparison, as the only possible sources of discrepancy are differences in the codes' algorithms.
In general, the \emph{FLASH} SZP1* simulations reach higher (potentially unstable) $CR$ values than \emph{MACH2} simulations, and \emph{MACH2} simulations reach higher temperatures than all \emph{FLASH} simulations.
The discrepant results highlight the sensitivity of the SZP1* configuration to heat transport processes (i.e., thermal conduction and radiation).
The high $CR$ values are the result of significant fuel thermal conduction losses.
As previously discussed, the SZP1* concept would benefit from decreasing thermal conductivity via fuel magnetization,
as was shown (perhaps erroneously) in the \emph{MACH2} model.
Such fuel magnetization could be achieved experimentally by applying an axial magnetic field to the configuration.

Despite the different results, all SZP1* simulations with both \emph{FLASH} and \emph{MACH2} generally agree on reaching peak fuel ion temperatures above 15 keV.
The highest-fidelity run, the \emph{FLASH} multi-group diffusion model, reaches the lowest peak ion temperature (see Fig.~\ref{fig:SZP1TionvsCR}), which in turn shows the importance of accurate radiation transport modeling for SZP1*.
At these observed high temperatures, alpha particle heating could be significant, but this physics capability is not available in \emph{FLASH} so we did not explore it with \emph{MACH2} either.
Any additional heat source or insulation, or increasing the initial fuel density, would help stagnate the fuel at a lower $CR$ value, thus improving stability.
It should be emphasized that \emph{experiments} of other SZP configurations, at smaller-than-Z pulsed-power facilities, have proven to be stable, and SZP1* is a theoretical platform in a different regime that may be more difficult to stabilize.

We eventually want to use \emph{FLASH} to simulate the entire spatial and temporal evolution of the SZP with a reactor-level drive current in three dimensions, taking full advantage of the extended-MHD and transport capabilities of the code.
The next immediate step is to conduct two-dimensional simulations of the models discussed in this work and in previous publications.\cite{Ruskov2020,Ruskov2021b}
Future work will assess the stability of the pinch (liner and target) to MHD instabilities in the presence/absence of axial magnetic fields, and explore how \emph{FLASH}'s extended-MHD terms can affect implosion dynamics and plasma conditions at stagnation.
This will shed light on the importance of previously unexplored physical processes at play in the SZP concept and contribute to the evaluation of the feasibility of the concept to achieve fusion.

\begin{acknowledgments}
This material is based upon work supported by the Department of Energy (DOE) National Nuclear Security Administration (NNSA) under Award Numbers DE-NA0003856, DE-NA0003842, DE-NA0004144, and DE-NA0004147, under subcontracts no. 536203 and 630138 with Los Alamos National Laboratory, and under subcontract B632670 with Lawrence Livermore National Laboratory. We acknowledge support from the U.S. DOE Advanced Research Projects Agency-Energy (ARPA-E) under Award Number DE-AR0001272 and the U.S. DOE Office of Science under Award Number DE-SC0023246. This work used resources of the National Energy Research Scientific Computing Center (NERSC), a U.S. DOE Office of Science User Facility located at Lawrence Berkeley National Laboratory, operated under Contract No. DE-AC02-05CH11231 using NERSC award ERCAP0024319. 
The authors acknowledge support from the High-Performance Computing group of the Laboratory for Laser Energetics at the University of Rochester.
The software used in this work was developed in part by the U.S. DOE NNSA- and U.S. DOE Office of Science-supported Flash Center for Computational Science at the University of Chicago and the University of Rochester.
This paper describes objective technical results and analysis.
Any subjective views or opinions that might be expressed in the paper do not necessarily represent the views of the U.S. DOE or the United States Government.
\end{acknowledgments}

\bibliography{mylibrary}

\end{document}